\documentclass[11pt]{article}
\usepackage{colortbl}
\usepackage{natbib}
\usepackage{verbatim}
\usepackage{amssymb,amsfonts,amsmath,amsthm}
\usepackage{geometry}
\usepackage{alltt}
\usepackage{multirow}
\usepackage{graphicx}
\usepackage{subfigure}
\usepackage{setspace}
\usepackage{appendix}
\usepackage[colorlinks,bookmarksopen,bookmarksnumbered,citecolor=blue,urlcolor=green,hypertexnames=false]{hyperref}
\usepackage{dsfont}
\usepackage{geometry}
\usepackage{animate}
\bibpunct[ ]{(}{)}{;}{a}{}{,}
\onehalfspace
\begin{document}
\title{Spatio-temporal adaptive penalized splines with application to Neuroscience\footnote{This paper is based on that published by the same authors as a part of the proceedings of the 31st International Workshop on Statistical Modelling, INSA Rennes, 4--8 July 2016 (Volume I, pp. 267\,--\,272. Eds: Dupuy, J-F. and Josse, J.)}}%
\author{Mar\'ia Xos\'e Rodr\'iguez - \'Alvarez$^{1,2}$, Mar\'ia Durb\'an$^{3}$, Dae-Jin Lee$^{1}$,\\Paul H. C. Eilers$^{4}$, Gonzalez, F.$^{5}$\\\\       
				\small{$^{1}$ BCAM - Basque Center for Applied Mathematics, Bilbao, Spain}\\
				\href{mailto:mxrodriguez@bcamath.org}{mxrodriguez@bcamath.org}\\
				\small{$^{2}$ IIKERBASQUE, Basque Foundation for Science, Bilbao, Spain}\\
        \small{$^{3}$ Department of Statistics, Universidad Carlos III de Madrid, Legan\'es, Spain}\\
        \small{$^{4}$ Erasmus University Medical Centre, Rotterdam, the Netherlands}\\                
		\small{$^{5}$ Department of Surgery and CIMUS, University of Santiago de Compostela, Spain}}
\maketitle
\date{}
\sloppy
\begin{abstract}
Data analysed here derive from experiments conducted to study neurons' activity in the visual cortex of behaving monkeys. We consider a spatio-temporal adaptive penalized spline (P-spline) approach for modelling the firing rate of visual neurons. To the best of our knowledge, this is the first attempt in the statistical literature for locally adaptive smoothing in three dimensions. Estimation is based on the \textit{Separation of Overlapping Penalties} (SOP) algorithm, which provides the stability and speed we look for.

\end{abstract}
\section{Visual receptive fields}
Electrophysiology studies record the electrical activity produced by neurons. They allow the study of the association between sensory stimuli and neural response in any part of the brain. Neurons produce sudden changes in their membrane potential known as `spikes', that can be recorded using microelectrodes. The analysis of the frequency of spike discharges provides insights on how the neurons and the nervous system work. 

Visual receptive fields (RFs) are small areas of the visual field that a particular visual neuron `sees'. Reverse cross-correlation is a receptive field mapping technique used for studying how visual neurons process signals from different positions in their receptive field. From the neuron responses (spikes) we can infer the spatio-temporal properties of the RFs (i.e., when and where a sensory stimulus produces a response). A detailed explanation on how the reverse cross-correlation technique was used in the experiments analyzed here can be found elsewhere \cite{Rodriguez2012}. Schematically, the subject (a monkey) was viewing two monitors (one for each eye) with a fixation target. Within a square area a bright or dark spot was flashed at different positions in a pseudorandom manner. Neuron spikes were recorded while the stimulus was delivered. When a spike was produced, the stimulus position at several pre-spike times was read. As a result, a set of numerical matrices (one for each pre-spike time) containing the number (counts) of times the stimulus was at that given position when a spike occurred is obtained. The graphical representation of each of these matrices is called receptive field map (RFmap), and can be regarded as a representation of the firing rate of the neuron.

\section{Three-dimensional adaptive P-spline}\label{MX:AdaptP_3D}
For each neuron, the reverse cross-correlation technique provides a dataset consisting of a series of $16$ matrices of dimension $16\times 16$, each matrix corresponding to the different pre-spike times considered (between $-20$ to $-320$ milliseconds). We adopted a Poisson model which expresses the neuron response (i.e., number of spikes) as a smooth function of both space and time 
\begin{equation}
log\left(E\left[y \mid r,c,t\right]\right) = log\left(n_{rc}\lambda_{rct}\right) = log\left(n_{rc}\right) + f\left(r,c,t\right),
\label{MX:PoissonModel}
\end{equation}
where $r$ indicates the row of the matrix, $c$ the column ($r,c = 1, \ldots, 16$), and $t$ the pre-spike time ($t = -20, \ldots, -320$). $n_{rc}$ denotes the number of stimulus presentations on each particular grid position of the square area (the offset) and $\lambda_{rct}$ is the intensity parameter (or firing rate). The smooth function $f(\cdot, \cdot, \cdot)$ was represented by the tensor product of three univariate B-spline basis \citep{Eilers2003}, i.e., $f(\boldsymbol{r}, \boldsymbol{c}, \boldsymbol{t}) = \left(\boldsymbol{B}_3^{(16 \times c_3)}\otimes\boldsymbol{B}_2^{(16 \times c_2)}\otimes\boldsymbol{B}_1^{(16 \times c_1)}\right)\boldsymbol{\theta}$, where $\otimes$ denotes the Kronecker product. 

In order to avoid over-fitting, the previous model can be estimated by penalized-likelihood methods \citep{Eilers2003}. In the absence of locally adaptive smoothness, the anisotropic penalty matrix is defined as
\begin{equation}
\lambda_{1}\left(\mathbf{I}_{c_3}\otimes\mathbf{I}_{c_2}\otimes\boldsymbol{D}_{1}^{t}\boldsymbol{D}_{1}\right) + \lambda_{2}\left(\mathbf{I}_{c_3}\otimes\boldsymbol{D}_{2}^{t}\boldsymbol{D}_{2}\otimes\mathbf{I}_{c_1}\right) + \lambda_{3}\left(\boldsymbol{D}_{3}^{t}\boldsymbol{D}_{3}\otimes\mathbf{I}_{c_2}\otimes\mathbf{I}_{c_1}\right),
\label{MX:2DPenalty}
\end{equation}
where $\lambda_1$, $\lambda_2$ and $\lambda_3$ are the smoothing parameters, and $\boldsymbol{D}_{d}$ ($d = 1,\;2,\;3$) are difference matrices of possibly different order $q_d$. 

In adaptive P-spline smoothing \citep[see, e.g.,][]{Rodriguez2015b} each $\lambda_{d}$ in (\ref{MX:2DPenalty}) is replaced by a vector of smoothing parameters $\boldsymbol{\lambda}_{d}$, where each component is associated with one coefficient difference (along the $d$-direction). However, this approach would imply as many smoothing parameters as coefficient differences, which could lead to under-smoothing and unstable computations. To reduce the dimension, $\boldsymbol{\lambda}_{d}$ is modelled by means of B-splines, i.e., 
\begin{align*}
\boldsymbol{\lambda}_{1} & = \left(\boldsymbol{C}_{11}^{((c_1 - q_1)\times p_{11})}\otimes\boldsymbol{C}_{12}^{(c_2\times p_{12})}\otimes\boldsymbol{C}_{13}^{(c_3\times p_{13})}\right)\boldsymbol{\phi}_1 = \boldsymbol{C}_{1}\boldsymbol{\phi}_1,\\
\boldsymbol{\lambda}_{2} & = \left(\boldsymbol{C}_{21}^{(c_1\times p_{21})}\otimes\boldsymbol{C}_{22}^{((c_2 - q_2)\times p_{22})}\otimes\boldsymbol{C}_{23}^{(c_3 - p_{23})}\right)\boldsymbol{\phi}_2 = \boldsymbol{C}_{2}\boldsymbol{\phi}_2,\\
\boldsymbol{\lambda}_{3} & = \left(\boldsymbol{C}_{31}^{(c_1\times p_{31})}\otimes\boldsymbol{C}_{32}^{(c_2\times p_{32})}\otimes\boldsymbol{C}_{33}^{((c_3 - q_3)\times p_{33})}\right)\boldsymbol{\phi}_3 = \boldsymbol{C}_{3}\boldsymbol{\phi}_3,
\end{align*}
where $\boldsymbol{C}_{ij}$ ($i,j = 1,\;2,\;3$) are B-spline regression matrices, with less columns than rows to ensure that the dimension is in fact reduced. The \textit{adaptive} penalty matrix in three dimensions can be then expressed as
\begin{align}
& \sum_{s = 1}^{p_{11}p_{12}p_{13}}\phi_{1s}\left(\mathbf{I}_{c_3}\otimes\mathbf{I}_{c_2}\otimes\boldsymbol{D}_{1}\right)^{t}diag\left(\boldsymbol{c}_{1,s}\right)\left(\mathbf{I}_{c_3}\otimes\mathbf{I}_{c_2}\otimes\boldsymbol{D}_{1}\right) & + \nonumber \\
& \sum_{u = 1}^{p_{21}p_{22}p_{23}}\phi_{2u}\left(\mathbf{I}_{c_3}\otimes\boldsymbol{D}_{2}\otimes\mathbf{I}_{c_1}\right)^{t}diag\left(\boldsymbol{c}_{2,u}\right)\left(\mathbf{I}_{c_3}\otimes\boldsymbol{D}_{2}\otimes\mathbf{I}_{c_1}\right) & + \label{MX:3DAP}\\
& \sum_{v = 1}^{p_{31}p_{32}p_{33}}\phi_{3v}\left(\boldsymbol{D}_{3}\otimes\mathbf{I}_{c_2}\otimes\mathbf{I}_{c_1}\right)^{t}diag\left(\boldsymbol{c}_{3,v}\right)\left(\boldsymbol{D}_{3}\otimes\mathbf{I}_{c_2}\otimes\mathbf{I}_{c_1}\right)\nonumber,
\end{align}
where $\boldsymbol{c}_{d,l}$ denotes the column $l$ of $\boldsymbol{C}_d$.

Estimation of the three-dimensional P-spline model for Poisson data (\ref{MX:PoissonModel}) subject to the adaptive penalty defined in (\ref{MX:3DAP}) can be based on its mixed-model representation. Restricted maximum likelihood (REML) estimates of the variance components (or smoothing parameters) are obtained by means of the \textit{Separation of Overlapping Penalties} (SOP) algorithm, recently proposed by \cite{Rodriguez2015a,Rodriguez2015b}. It should be noted that the reformulation of model (\ref{MX:PoissonModel}) as a mixed model does not gives rise to a diagonal precision matrix, and thus, some of the computational advantages of SOP are lost. Nevertheless, even in this case, the algorithm provides reasonable computing times. Besides, Generalized Linear Array Models \citep[GLAM,][]{Currie2006} can be used to compute the inner products involved in the mixed model equations as well as the penalty matrices given in (\ref{MX:3DAP}), thus improving the speed of the estimation algorithm. 
  
\section{Results} 
\begin{figure}[ht!]\centering
\includegraphics[width=15cm]{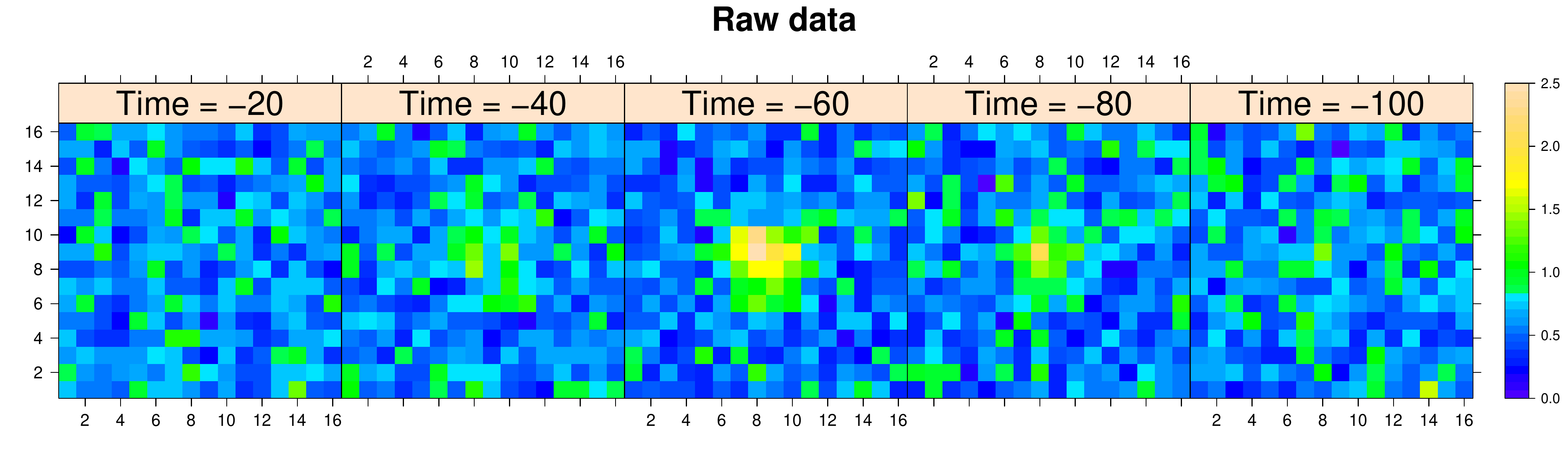}
\includegraphics[width=15cm]{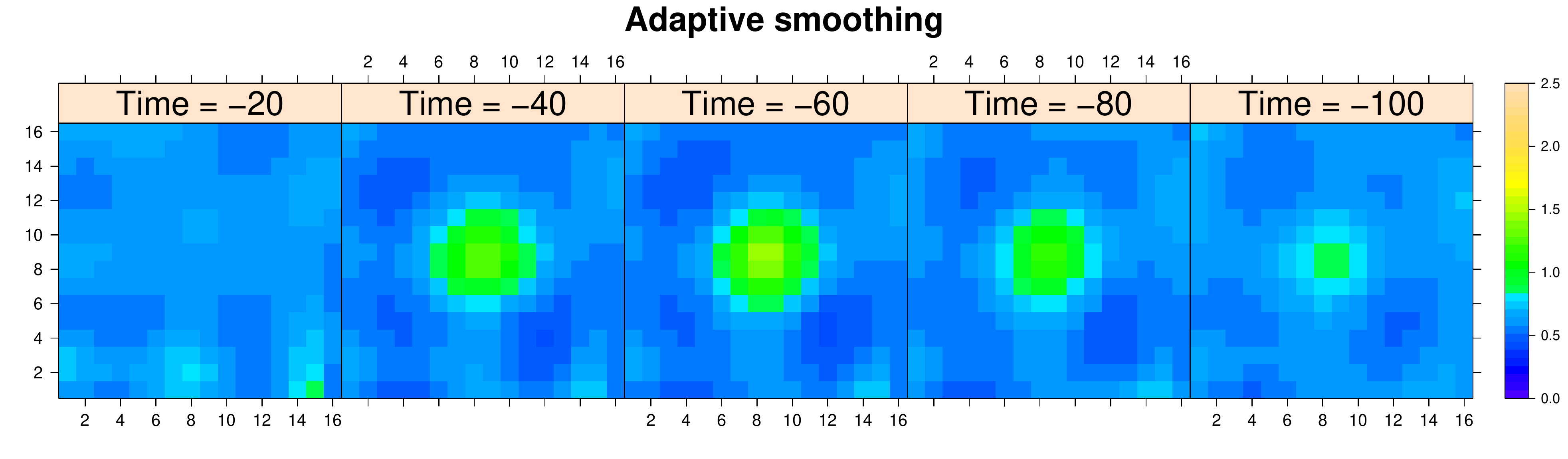}
\includegraphics[width=15cm]{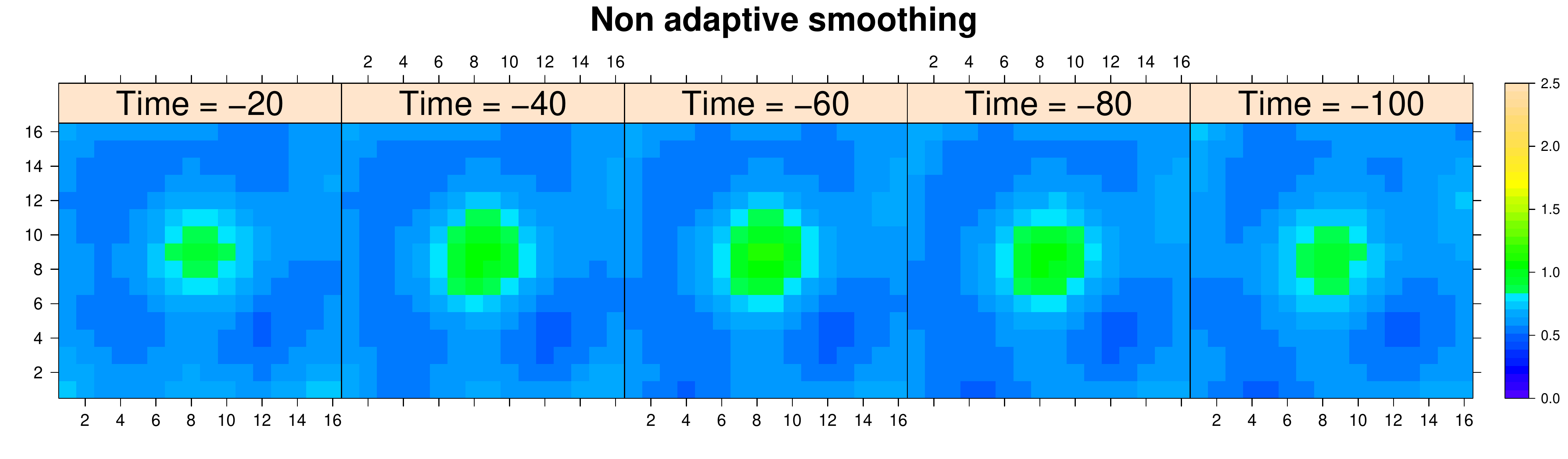}
\caption{\label{MX:FAP1} Level plot of the observed and smoothed firing rates of the RFmap with and without locally adaptive smoothness.}
\end{figure}

\begin{figure}[ht!]\centering
\includegraphics[width=15cm]{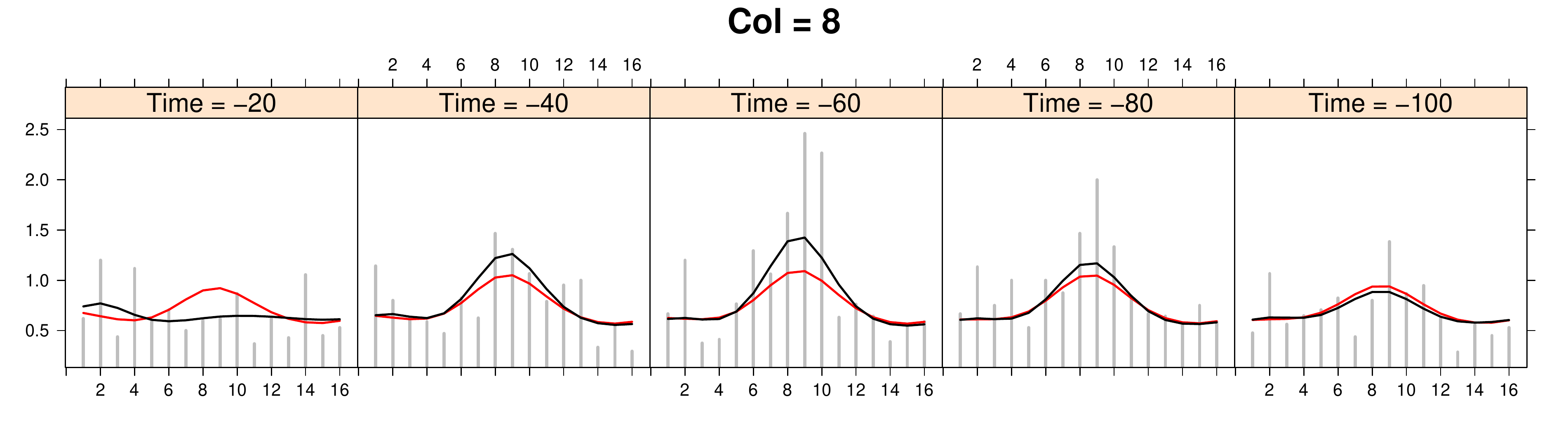}
\includegraphics[width=15cm]{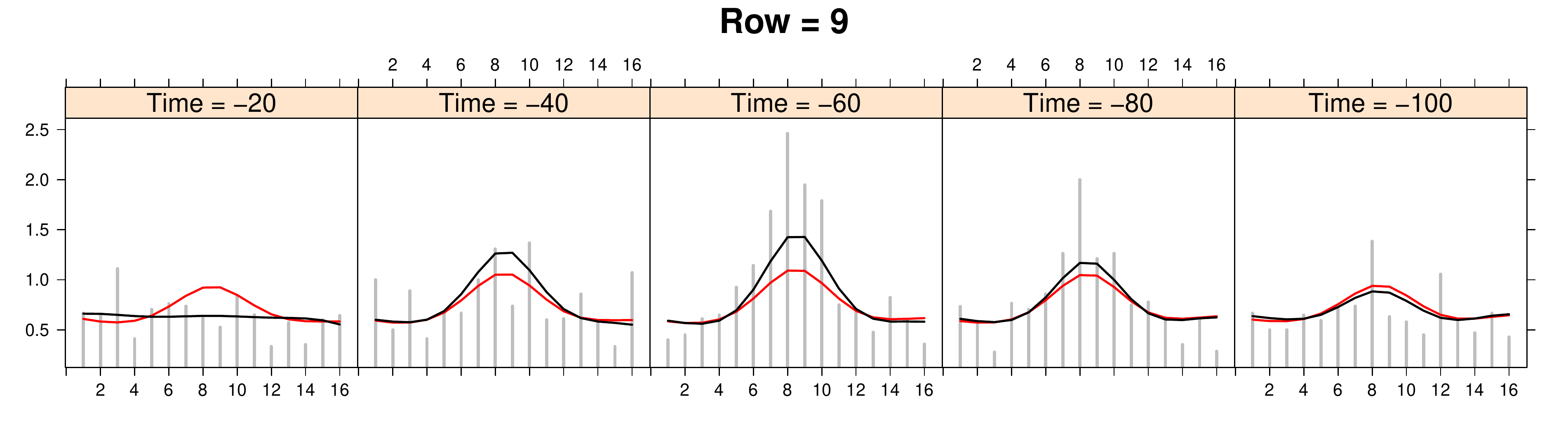}
\caption{\label{MX:FAP2} Observed and smoothed firing rates of the RFmap by row for column 8 (top figure); and by column for row 9 (bottom figure). Gray vertical lines: observed. Black line: adaptive approach. Red line: non adaptive approach.}
\end{figure}

For illustration purposes, we show the results for a single visual neuron from area V1 (primary visual cortical area). Model (\ref{MX:PoissonModel}) was estimated with and without assuming locally adaptive smoothness by means of the SOP algorithm and GLAM	. In both cases, we used second-order differences ($q_d = 2$) and marginal B-splines bases of dimension $c_d = 7$. For the adaptive approach, we chose $p_{ij} = 4$ (i, j = 1,2,3), yielding a total of $192$ ($3\times 4^3$) smoothing parameters (or variance components). Animation \ref{MX:FAP0} and Figure \ref{MX:FAP1} show the observed and estimated series of smooth RFmaps for several pre-spike times using both approaches. As it can be seen, both analyses show a central area of high values that represents the visual RF of the neuron, which is in concordance with the raw data. However, there are two major differences: whereas the non adaptive approach seems to indicate that the time between sensory stimulus and response spans from $20$ to $100$ ms, the adaptive method reduces this time span from $40$ to $100$ ms. Also the adaptive approach shows a sharper increase and a larger estimate of the firing rate than the non-adaptive approach (see also Figure \ref{MX:FAP2}). In terms of computational effort, in the absence of adaptive smoothness the algorithm needed about $14$ seconds, whereas the complexity afforded by the adaptive approach increased the computing time to $133$ seconds.

\begin{figure}
\centering
\animategraphics[autoplay,loop,width=15cm]{1}{AnimFull_f}{0}{15}
\caption{\label{MX:FAP0} Animation with the observed and smoothed firing rates of the RFmap with and without locally adaptive smoothness.}
\end{figure}
 
\section*{Acknowledgments}{The authors express their gratitude for the Spanish Ministry of Economy and Competitiveness MINECO grants MTM2014-55966-P, MTM2014-52184-P and BCAM Severo Ochoa excellence accreditation SEV-2013-0323, and for the Basque Government grant BERC 360 2014-2017. This study was also partially supported by Plan Nacional 2013-2016, ISCIII (RETICS, Oftared) RD12/0034/0017, and FEDER.}


\begin{thebibliography}{200}
\bibitem[Currie et al., 2006]{Currie2006}Currie, I., Durb\'an, M. and Eilers, P.H.C.] (2006). Generalized linear array models with applications to multidimensional smoothing. {\it Journal of the Royal Statistical Society, Series B}, \textbf{68}, 259\,--\,280.
\bibitem[Eilers and Marx (2003)]{Eilers2003}Eilers, P.H.C. and Marx, B.D. (2003). Multivariate calibration with temperature interaction using two-dimensional penalized signal regression. {\it Chemometrics and intelligent laboratory systems}, {\bf 66}, 159\,--\,174.
\bibitem[Rodr\'iguez-\'Alvarez et al. (2012)]{Rodriguez2012} Rodr\'iguez-\'Alvarez, M.X., Cadarso-Su\'arez, C., Gonz\'alez, F. (2012). Analysing Visual Receptive Fields through Generalised Additive Models with Interactions. {\it SORT}, {\bf 36}, 3\,--\,32.
\bibitem[Rodr\'iguez-\'Alvarez et al. (2015a)]{Rodriguez2015a} Rodr\'iguez-\'Alvarez, M.X., Lee, D-J., Kneib, T., Durb\'an, M. and Eilers, P.H.C. (2015a). Fast smoothing parameter separation in multidimensional generalized P-splines: the SAP algorithm. {\it Statistics and Computing}, {\bf 25}, 941\,--\,957.  
\bibitem[Rodr\'iguez-\'Alvarez et al. (2015b)]{Rodriguez2015b} Rodr\'iguez-\'Alvarez, M.X., Durb\'an, M., Lee, D-J. and Eilers, P.H.C. (2015b). Fast estimation of multidimensional adaptive P-spline models.In: {\it Proceedings of the 30th Workshop on Statistical Modelling}, Friedl, H. and Wagner, H. (Eds. ), pp. 330\,--\,335.
\end{thebibliography}
\end{document}